\documentclass{aa}  

\usepackage{graphicx}
\usepackage{txfonts}
\usepackage{hyperref}
\usepackage{siunitx}
    \DeclareSIUnit{\astronomicalunit}{AU}
\usepackage{subcaption}
\usepackage{array}
	\newcolumntype{C}{>{$}c<{$}}

\usepackage{color}

\begin{document} 

    \bibliographystyle{aa}
    \title{The Kepler-223 resonance holds information on turbulence during the gas disk phase}
    \titlerunning{The Kepler-223 resonance holds information on its gas disk turbulence}

    \author{L.-A. H\"uhn\inst{1}
        \and
        G. Pichierri\inst{1}
        \and
        B. Bitsch\inst{1}
        \and
        K. Batygin\inst{2}
        }

    \institute{Max-Planck-Institut f\"ur Astronomie, K\"onigstuhl 17, 69117 Heidelberg\\
            \email{huehn@mpia.de}
        \and
            Division of Geological and Planetary Sciences, California Institute of Technology, 1200 E. California Blvd., %
            Pasadena, CA 91125, USA
            }

    \date{\today}
 
    \abstract
    {Planet formation remains an open field of research, and many fundamental physical processes regarding planetary formation in protoplanetary disks are still imperfectly understood. It remains to be investigated how different conditions in those protoplanetary disks affect how different kinds of observed systems emerge, with an elusive phenomenon being turbulence in such disks. Observations are available of planetary systems and of some protoplanetary disks, which can serve as a starting point for these investigations. Detected systems reveal different architectures of planets. One particularly interesting case to consider is the Kepler-223 system, which contains a rare configuration of four planets in a resonance chain, implying a certain migration history.}
    {We aim to use the orbital configuration of Kepler-223's planets to constrain the parameters of the protoplanetary disk that allow for the formation of a chain of mean-motion resonances that resembles the one of Kepler-223. The parameters we aim to investigate are primarily the disk viscosity and surface density.}
    {We use the \texttt{swift\_symba} N-body integrator with additional dissipative forces to mimic planet-disk interactions.}
    {We constrain the surface densities and viscosities that allow the formation of a resonant chain like that of Kepler-223. We find that surface densities of up to a few minimum mass solar nebula (MMSN) surface densities and disk viscosity parameters $\alpha$ of a few \num{e-3} up to \num{e-2} are most successful at reproducing the architecture of this particular planetary system. We describe how these two quantities are linked to each other considering the success of reproducing the chain and find that higher disk surface densities in turn require lower viscosities to build the chain.}
    {Our results show that well characterized observed planetary systems hold information about their formation conditions in the protoplanetary disks and that it is possible to extract this information, namely the initial disk surface density and viscosity, helping to constrain planet formation.}


    \maketitle
%

\section{Introduction}

 The study of exoplanetary systems is a key element for understanding the formation and evolution of planets, as well as the assembly of planetary systems and their dynamical evolution. Currently, the exoplanet sample is largely dominated by data from the Kepler mission, which provided well constrained planetary radii and orbital periods using the transit detection method. Out of all detected planets, small and close-in planets, the so-called Super-Earths and Mini-Neptunes, are thought to be orbiting up to 50\% of sun-like stars \citep{howard2012,mayor2011,mulders2018}, and they often occur in multiplanetary configurations, where the planets within each system mostly have similar masses and radii (e.g., \citealt{millholland2017,weiss2018}).
 
 A notable example is the planetary system hosted by Kepler-223, a G5V star with a mass of $1.1\mathrm{M}_\sun$. Four Sub-Neptunian planets in mean motion resonance (MMR) are orbiting it \citep{mills2016}. In addition to orbital periods and radii, \citet{mills2016} used transit timing variations (TTVs) to obtain masses, eccentricities and inclinations of the planets. The masses are given by $7.4_{-1.1}^{+1.3} M_\mathrm{Earth}$, $5.1_{-1.1}^{+1.7} M_\mathrm{Earth}$, $8.0_{-1.3}^{+1.5} M_\mathrm{Earth}$ and $4.8_{-1.2}^{+1.4} M_\mathrm{Earth}$, for planets b-e, respectively. The period ratios of the planets are 8:6:4:3 starting from the inner-most planet Kepler-223 b. This suggests the existence of individual MMRs of the pairs of planets in the MMR chain, that is a 4:3 resonance for the first pair, a 3:2 resonance for the second pair and a 4:3 resonance for the third pair. \citet{mills2016} confirmed the resonant nature of the system, showing librating Laplace angles. More detailed information about the planets orbiting Kepler-223 is presented in Table \ref{tab:kepler-223}.
 
\begin{table*}[htp]
\caption[Parameters of the planets orbiting Kepler-223]{Parameters of the planets orbiting Kepler-223 \citep{mills2016}.}
\label{tab:kepler-223}
\centering
\begin{tabular}{l|C*{6}{@{\hspace*{-0.3em}}C}}
    \hline\hline
    \textbf{Parameter} & \textbf{Kepler-223 b} && \textbf{Kepler-223 c} && \textbf{Kepler-223 d} && \textbf{Kepler-223 e}\\
    \hline
    \textbf{Eccentricity} & 0.078^{+0.015}_{-0.017} && 0.150^{+0.019}_{-0.051} && 0.037^{+0.018}_{-0.017} && 0.051^{+0.019}_{-0.019}\\
    \textbf{Mass }($M_\mathrm{Earth}$) & 7.4^{+1.3}_{-1.1} && 5.1^{+1.7}_{-1.1} && 8.0^{+1.5}_{-1.3} && 4.8^{+1.4}_{-1.2}\\
    \textbf{Orbital period }(d) & 7.38449^{+0.00022}_{-0.00022} && 9.84564^{+0.00052}_{-0.00051} && 14.78869^{+0.00030}_{-0.00027} && 19.72567^{+0.00055}_{-0.00054}\\
    \textbf{Period ratio} && \text{4:3} && \text{3:2} && \text{4:3} &\\
    \hline
\end{tabular}
\end{table*}%

 While capturing into a MMR is a natural outcome of type-I migration, they are found very infrequently in the exoplanetary sample. It was suggested that capture into a resonance chain does not produce dynamically stable systems often, if the disk in inviscid \citep{mcnally2019}. Another often studied reason for the rarity of MMRs is their instability during (e.g.,\ \citealt{liu2017}) or after the dispersal of the protoplanetary disk (e.g.,\ \citealt{izidoro2017,izidoro2019,pichierri2020,lambrechts2019}).  Thus, the existence of the particular configuration of the Kepler-223 system, as well as other resonant chains like Trappist-1 \citep{gillon2016,gillon2017,luger2017} or GJ-876 \citep{marcy2001,rivera2005}, poses a definite challenge to formation scenarios of planetary systems. For Kepler-223, the resonant state appears to be long-term stable (it is, for example, protected from instabilities due to the emergence from secondary resonances \citep{pichierri2020}), raising the question of what condition brought about this particular configuration. Indeed, these exotic orbital configurations offer an unique opportunity, as the current state can be used to constrain parameters of models that describe the protoplanetary disk in which they formed.
 
 Important disk parameters that could be constrained this way are, for example, the surface density or a description of turbulent viscosity in the disk. The surface density is a measure for the mass, which in turn is an important parameter for the migration speed \citep{kley2012,baruteau2014}, and also relevant for example for gravitational instabilities. Turbulence in disks can arise from different hydrodynamical instabilities, for instance the global baroclinic instability \citep{klahr2003} or the vertical shear instability \citep{richard2016,flock2017,manger2020}, or from the magneto-rotational instability (MRI) \citep{balbus1991}. Turbulent viscosity is a key element of many planet formation scenarios and disk evolution models, which is solidified by the Atacama Large Millimeter/submillimeter Array (ALMA) data of disks that might show still ongoing planetary formation \citep{teague2018,pinte2018,keppler2018}, and is typically described in terms of the $\alpha$ parameter \citep{shakura1973}. Despite its significance and recent improvements, the $\alpha$ parameter is not very well constrained observationally, with values typically ranging between $\num{e-4}$ and about $\num{4e-2}$, as studies of disks observable by ALMA have shown \citep{rafikov2017,dullemond2018,flaherty2017,flaherty2018}. These values are also consistent with hydrodynamical and MRI simulations (e.g., \citealt{turner2014}).
 
 In this work, the fact that turbulence causes stochastic fluctuations of the disks density structure, leading to stochastic forces on the planets, is used to study its effects on the formation of Kepler-223's resonance chain. The main idea is that turbulence can be a reason for destroying or preventing the formation of a MMR chain even before the disk has dissipated (e.g.,\ \citealt{batygin2017}). The aim of this investigation is to constrain the $\alpha$ parameter in combination with other disk parameters, just from observational information contained in the current state of the planetary system. 
 
 The methods used to investigate Kepler-223 are laid out in Sect. \ref{sec:methods}. Kepler-223 serves as a test case for the methodology that we use to find constraints on the disk turbulence and surface density. However, the methods we apply are general and can be applied to other systems with resonant chains. The results of our investigation are presented in Sect. \ref{sec:results}. We discuss implications and shortcomings of our investigation in Sect. \ref{sec:discussion}.


\section{Methods}\label{sec:methods}

We reproduced the Kepler-223 resonance chain using N-body simulations using the \texttt{swift\_simba} code, with additional dissipative forces implemented to mimic the interactions of the planets with the protoplanetary disk, including the effects of stochastic surface density changes caused by turbulence.

\subsection{Type-I migration}
As a model for the effects on the semi-major axis and the eccentricity resulting from planet-disk interactions, the prescription developed by \citet{cresswell2008} was implemented. The effects of eccentricity damping and type-I migration are viewed as two separate contributions. The former is described by
\begin{equation}
    \dot{e}_\mathrm{damp} = -\frac{e}{\tau_e}\text{,}
\end{equation}
with the timescale $\tau_e$ given by
\begin{equation}
    \begin{split}
    \tau_e = \frac{\tau_\mathrm{wave}}{\num{0.780}}\left[1-0.14\left(\frac{e}{H/r}\right)^2+0.06\left(\frac{e}{H/r}\right)^3\right.\\
    \left.+0.18\left(\frac{e}{H/r}\right)\left(\frac{i}{H/r}\right)^2\right]\text{,}
    \end{split}
\end{equation}
Here, $\tau_\mathrm{wave}$ is the typical type-I migration timescale,
\begin{equation}
    \tau_\mathrm{wave} = \frac{h^4}{q}\frac{M_\star}{\Sigma_p a^2}\Omega_p^{-1}\text{,}\label{eq:tauwave}
\end{equation}
where $q=\frac{m}{M_\star}$ is the fraction of the planetary mass to the stellar mass, $h=H/r$ is the disk's aspect ratio and the subscript $p$ denotes that the quantity is evaluated at the planets location. Planetary type-I migration is modeled given the negative torque,
\begin{equation}
    \dot{\mathcal{L}} = -\frac{\mathcal{L}}{\tau_\mathrm{mig}}\text{,}
\end{equation}
with a migration timescale of
\begin{equation}
    \tau_\mathrm{mig} \approx 2\frac{\tau_\mathrm{wave}}{2.7+1.1\alpha_\Sigma}h^{-2}\text{,}
\end{equation}
again in the limit of small eccentricities, with the power law of the surface density $\alpha_\Sigma$ (see below). The resulting change of the semi-major axis can then be calculated as
\begin{equation}
    \frac{\dot{a}}{a} = -\frac{1}{\tau_a}-p\frac{e^2}{\tau_e}\text{,}
\end{equation}
with $\tau_a=\tau_\mathrm{mig}/2$ and $p\approx 2$ for small $e$. Note that the migration speed scales linearly with $q$ and the surface density $\Sigma$. The ratio of the migration to eccentricity damping time-scale, the $K$-factor \citep{ramos2017} is given by
\begin{equation}
    K=\frac{\tau_a}{\tau_e}=\frac{\tau_\mathrm{mig}}{2\tau_e}\approx\frac{0.780}{2.7+1.1\alpha_\Sigma}h^{-2}\text{.}\label{eq:kfactor}
\end{equation}
Equation \eqref{eq:kfactor} shows that, for typical values for $h$ and $\alpha_\Sigma$, eccentricity damping happens on a much shorter timescale than migration. Therefore, in this regime the planets are expected to have vanishing eccentricities before capturing into resonance.

To stop migration at the inner edge of the protoplanetary disk, a planetary trap is implemented. To model the sign flip of the torque and its radial dependence, which ultimately stops planetary migration once the planet reaches the inner edge \citep{masset2006,flock2019}, a multiplier $\mathcal{F}$ is applied to the acceleration of the planets \citep{pichierri2018}. It is given by
\begin{equation}
    \mathcal{F} = 
     \begin{cases}
       1\text{,} &\quad\text{if }a\geq r_e(1+h_e)\text{,}\\
       5.5\sin\left(\frac{(a-r_e)\pi}{2r_eh_e}\right)-4.5\text{,} &\quad\text{if }r_e(1-h_e)<a<r_e(1+h_e)\text{,}\\
       -10\text{,} &\quad\text{if } a\leq r_e(1-h_e)\text{.}\\
     \end{cases}
\end{equation}
Here, $r_e$ is the location of the edge (in \si{\astronomicalunit}) and $h_e$ is the aspect ratio of the disk at $r_e$. This implementation does not affect eccentricity damping.

\subsection{Disk-driven turbulence}
Our prescription for disk turbulence follows an implementation of \citet{ogihara2007} based on results from hydrodynamical simulations investigating MRI turbulence \citep{laughlin2004}. Turbulent modes are generated with a turbulent strength parameter $\gamma$. We summarize this method in the following. The stochastic force exerted by the turbulent disk is described by
\begin{equation}
    F_\mathrm{turb} = -\Gamma\nabla\Phi\text{,}
\end{equation}
where the parameter $\Gamma$ is given by
\begin{equation}
    \Gamma=\frac{64\Sigma r^2}{\pi M_\star}\text.
\end{equation}
The potential is described by a superposition of individual modes,
\begin{align}
    \Phi &= \gamma r^2\Omega^2\sum_{i=1}^{n_\mathrm{modes}}\Lambda_{c,m}\text{,}\\
    \Lambda_{c,m} &= \xi e^{\frac{(r-r_c)^2}{\sigma^2}}\cos(m\theta-\phi_c-\Omega_c\tilde{t})\sin\left(\pi\frac{\tilde{t}}{\Delta t}\right)\text{.}\label{eq:turb_modes}
\end{align}
In equation \eqref{eq:turb_modes}, $\xi$ describes a Gaussian distribution with $\sigma=1$ and $(r,\theta)$ the radial and azimuthal position of the planet, while $r_c$ and $\phi_c$ describe the center of the density fluctuations caused by the turbulent modes. The azimuthal extent of the modes is given by $\frac{2\pi r_c}{m}$, while the radial extent is given by $\sigma=\frac{\pi r_c}{4m}$. The time dependence of the modes is given by $\Delta t$. A mode is created at time $t_0$ and dies when $\Delta t>\tilde{t} \coloneqq t-t_0$. For $\Delta t$, the sound crossing time in angular direction ($\Delta t = \frac{2\pi r_c}{mc_s}$) was chosen. To create a new mode, $r_c$ is randomly generated inside the considered edges of our system $r_\mathrm{in}$ and $r_\mathrm{out}$ and $\phi_c\in[0,2\pi)$ is also chosen randomly. The wavenumber $m$ is chosen from a logarithmic random distribution in the range of $2\leq m \leq 64$. The resulting torque $\Gamma_t$ is then given by
\begin{equation}
    \Gamma_t=r\frac{1}{r}\frac{\partial\Phi}{\partial\theta}\Gamma m_\mathrm{pl} = -\gamma\Gamma m_\mathrm{pl}r^2\Omega^2\sum_{i=1}^{n_\mathrm{modes}}m\Lambda_{s,m}\text{,}\label{eq:kickstrength}
\end{equation}
where
\begin{equation}
    \Lambda_{s,m} = \xi e^{\frac{(r-r_c)^2}{\sigma^2}}\sin(m\theta-\phi_c-\Omega_c\tilde{t})\sin\left(\pi\frac{\tilde{t}}{\Delta t}\right)\text{.}
\end{equation}
Here, $m_\mathrm{pl}$ denotes the planetary mass in order to avoid confusion with the wavenumber $m$.

The stochastic force strength parameter $\gamma$ can be related to the $\alpha$ parameter typically used to describe turbulence in current disk models. To this end, we use on the one hand a relationship for the diffusion coefficient in the eccentricity $D_e$ to the model's strength parameter  \citep{ida2008,okuzumi2013}, where the numerical pre-factor used was modified to fit the setup used here (see appendix \ref{sec:de_fit}),
\begin{equation}
    D_e \approx \num{3e-4} \left(\frac{\gamma}{\num{e-3}}\right)^2 \left(\frac{\Sigma a^2}{M_\star}\right)^2 \Omega\text{;}\label{eq:de_gamma}
\end{equation}
on the other hand we use a similar relationship involving the $\alpha$ parameter \citep{okuzumi2013},
\begin{equation}
    D_e \approx \num{2.4e-3}\left(\frac{\alpha}{\num{e-2}}\right) \left(\frac{\Sigma a^2}{M_\star}\right)^2 \Omega\text{.}
\end{equation}
This yields a conversion between the two parameters that reads
\begin{equation}
    \gamma \approx \num{3e-3} \left(\frac{\alpha}{\num{e-2}}\right)^{1/2}\text{.}\label{eq:gammaalpha}
\end{equation}
This relationship should be considered valid as an order-of-magnitude correspondence between $\gamma$ (the parameter used in the disk model for generating turbulent waves) and $\alpha$ (the commonly used turbulence parameter).


\section{Results}\label{sec:results}

\subsection{Results without turbulence}
\subsubsection{Treatment of individual planet pairs}\label{sec:individual_pairs}

As a starting point, we treated all three planet pairs of the Kepler-223 system separately. We first simulate each resonant pair separately for varying surface density values $\Sigma=\Sigma_0 (r/r_0)^{-\alpha_\Sigma}$ in order to get a first overview over how the surface density distribution affects each individual pair. For all conducted simulations here and in later sections, the aspect ratio was fixed at $h=0.05$. The power law of the surface density $\alpha_\Sigma$ was kept constant at a value of $\alpha_\Sigma=1.5$. The reference distance $r_0$ was chosen to be $r_0=\SI{1}{\astronomicalunit}$. For the purposes of this step, we set the trap at the inner disk edge at the position of the inner planet in each pair under investigation. This allows the inner planet to be fixed allowing a clear investigation of the convergent migration behavior, and in turn means that $\Sigma_0$ sufficiently describes the surface density. In fact, the planets also capture in order, from the inner-most to the outer-most one, when considering the case of all four planets, so this is a good simplification to make to gain first insights into the migration behavior.

Without considering turbulence and for fixed planet parameters, the migration timescale is the quantity that mainly defines the migration behavior and speed, and thus the final resonance. It is a function of the surface density, which makes it an important quantity to investigate for migration. It is important to keep in mind that we do not actually know at which distance from the star the four Kepler-223 planets formed, nor when. Therefore, for the purposes of this initial step, the planets of each pair were put wide of the 2:1 MMR initially, only in order to find a rough first estimate for the values for $\Sigma_0$ that we expect to work for the formation of the specific resonance chain. As a general rule, for a higher surface density, resonances with lower $k$ are skipped, subsequently leading to a capture into a resonance with a higher $k$ \citep{batygin2015}. Therefore, we consider a range of values for $\Sigma_0$ for the individual pairs and check the index $k$ of the resonance the pairs capture into. Considering the different location and different masses of the pairs, we expect a value for $\Sigma_0$ to lead to different final $k$ for each individual pair. We then compare these final $k$ to the resonance indices measured by \citet{mills2016} to find a range of surface densities that allow capture into the correct resonance.

\begin{figure}
    \centering\includegraphics[width=.5\textwidth]{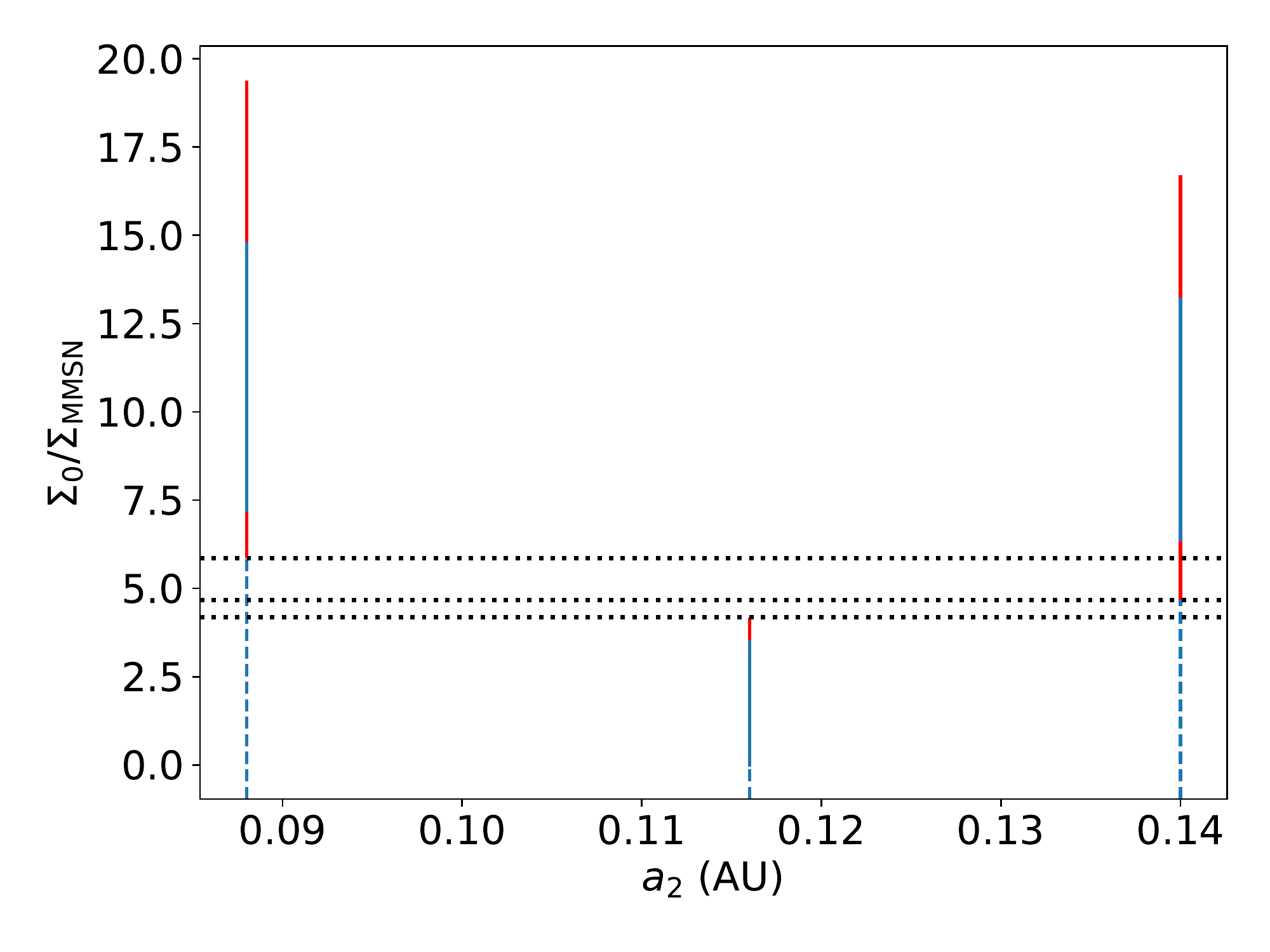}
    \caption{Range of surface density values $\Sigma_0$ that allow the individual Kepler-223 planet pairs to capture into the correct resonance, as a function of the semi-major axis of the outer planet of each pair. In red, the error is indicated, which is related to the mass error of the planets \citep{mills2016}. The black dotted lines are shown to make clear that within the errors, no value for $\Sigma_0$ can be found that allow each separate pair to capture into the correct resonance, if the pairs are initially wide of the 2:1 MMR. The blue dotted vertical lines indicate that this does not hold if the planets are instead placed just wide of the correct MMR, which eliminates the lower limit for $\Sigma_0$.}
    \label{fig:individual_pairs}
\end{figure}

The result of this consideration is shown in Fig. \ref{fig:individual_pairs}. Under the strong assumption that the planets formed outside of the 2:1 MMR, one cannot find a surface density power-law prescription that allows all pairs to individually capture into the correct resonance. While in principle other slope-values $\alpha_\Sigma$ than the used value of $3/2$ are possible, this statement is independent of this parameter. However, when placing the individual pairs in front of the desired resonances, there is no lower limit for the surface density that allows the planets to be captured into that resonance. In that case, one can indeed find surface density values that allow the formation of the observed resonance for all pairs individually. All in all, this investigation is of course insufficient for the problem at hand, and should rather be taken as an estimate for the relevant range of $\Sigma_0$ for a more complete consideration. Naturally when considering a resonant chain of four planets, four-planet dynamics need to taken into account, which also change the outcome with respect to the final configuration per surface density.

\subsubsection{Treatment of all four planets together}
The next natural step to investigate the capture into a resonant chain of the Kepler-223 system is thus to consider all four planets at once. We consider different surface densities, expecting it to influence the migration timescale, and different initial planet locations. For the planets' locations, we take a pragmatic approach. As we already mentioned their initial formation location can not be inferred from observations and could have been in principle in a configuration where individual planet pairs lie far outside the currently observed resonances. For the purpose of this investigation, however, the planets are placed just outside the resonances in which they are observed today. The reason for this is that, since the innermost planet is locked at the inner edge of the disk \citep{masset2006,ataiee2021}, migration is typically convergent anyway (see below), and thus if a planet pair is observed today in a certain mean motion resonance, it is natural to assume, within the framework of our investigation, that they once lied just wider of that specific period ratio. Note that this initial placement does not pose a contradiction to the results of Sect. \ref{sec:individual_pairs}. In that section, the premise was to find a range of potential surface density values that locally allow the formation of a resonance with index $k$ that matches the observations by \citet{mills2016}. This was achieved by considering the analytically well-understood case of a two-planet system without turbulence. In the investigation in this section, the dynamics of the four-planet system while including effects of turbulence are considered. In this scenario, planets are able to reach the position where we put them at the start of the simulation, even if they do form wider of the resonances they are presently observed in and have to migrate inward. The distance to an unperturbed semi-major axis ratio is parameterized by $\varepsilon$,
\begin{equation}
    \frac{a_{i+1}}{a_i} = \left(\frac{k_i+1}{k_i}(1+\varepsilon_i)\right)^{2/3}\text{,}
\end{equation}
where $i=1,2,3$ denotes the planet pair of the chain, and $k_i$ is the resonance index, with $k_1=4$, $k_2=3$ and $k_3=4$. The first planet's initial semi-major axis is set to $a_1=0.076\,\mathrm{AU}$, which is close to the current location of Kepler-223b given by Kepler's third law, $a_b=\SI{0.0772}{\astronomicalunit}$ \citep{mills2016}.\\

\begin{figure*}[htp]
    \centering\includegraphics[width=\textwidth]{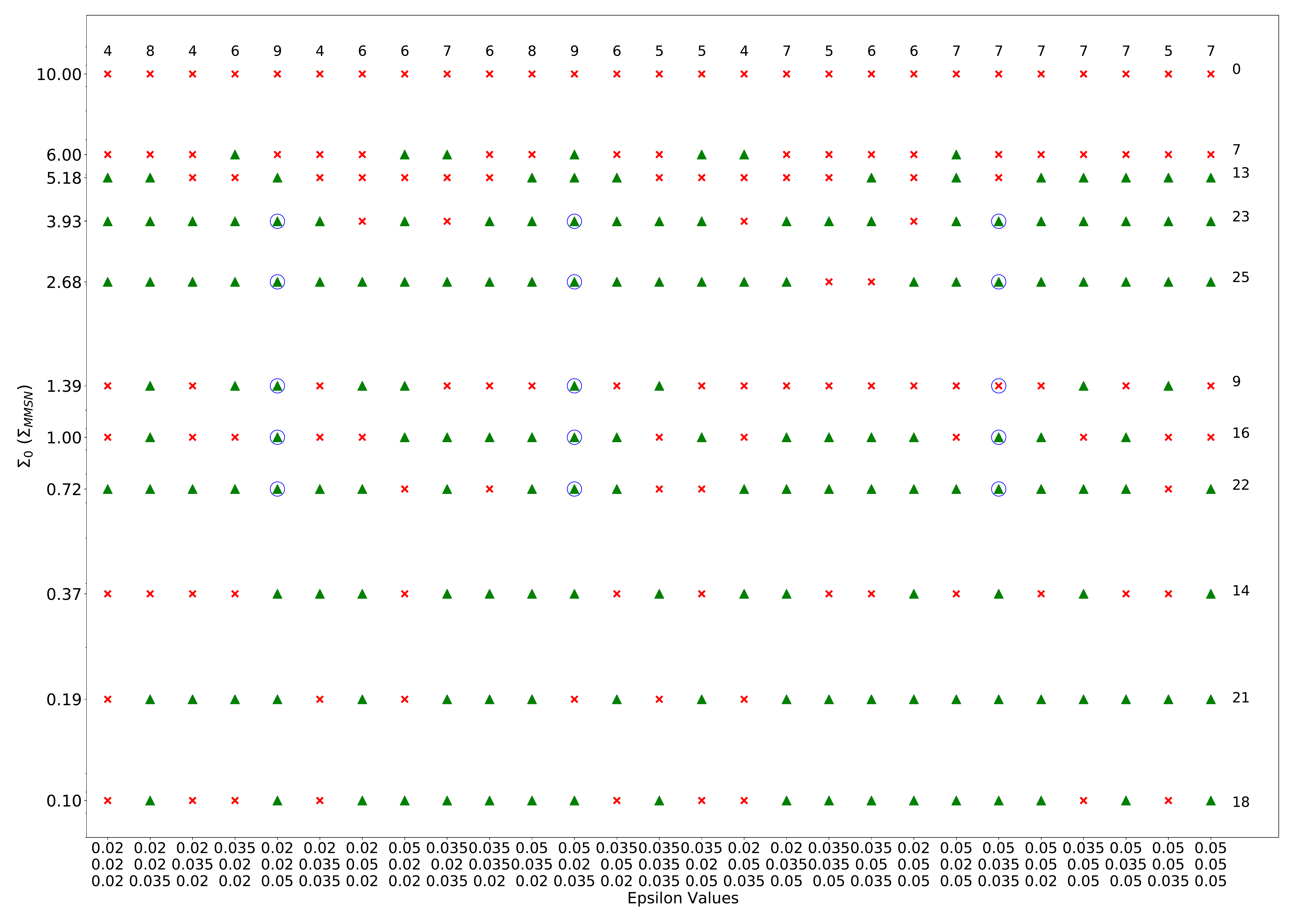}
    \caption{The buildup of the resonance chain of the four Kepler-223 planets, without considering turbulence. The investigated parameter space is two-dimensional: The surface density, as well as the initial separation of the planet pairs from the final resonance, are varied. Configurations where the correct chain was built are indicated by green triangles, while those who failed to do so are shown as red crosses. The vales for $\varepsilon$ are ordered such that $\varepsilon_1$ is the top-most value. Simulations whose symbol is encircled share their initial parameters with those considered for a consideration including disk turbulence (Sect. \ref{sec:turb}).}
    \label{fig:dotsplot}
\end{figure*}

We run N-body simulations for values of $\Sigma_0$ ranging between 0.1 and 10 $\Sigma_\mathrm{MMSN}$, for all combinations of values for $\varepsilon_i$ out of \{0.02,0.035,0.05\} representing small, medium and large initial separations between the planets. Even for the largest initial separation, the planet pair was always placed in such a way that no first-order MMR had to be crossed in order to reach the intended one. After the integration, a system was considered to exhibit the correct chain of resonances if their period ratios were wide of the observed resonance, as well as the corresponding resonant angles showing libration. A system's configuration of periods was considered to be compatible with observations if the ratio of the periods of each individual planet pair exceeded the nominal value, but with a relative error below $0.1\%$. The result of this investigation can be seen in Fig. \ref{fig:dotsplot}. One can see the surface density value at which the migration timescale is too short to be able to form the Kepler-223 resonance chain, because the needed resonances are skipped as expected. This happens beyond a surface density corresponding to ${\sim}6\,\Sigma_\mathrm{MMSN}$. This demonstrates at face value that, at least without any treatment of turbulence, the minimum-mass extrasolar nebula (MMEN, \citealt{chiang2013}), cannot accurately describe protoplanetary disks of extrasolar systems like Kepler-223. It predicts a surface density of ${\sim}7\Sigma_\mathrm{MMSN}$ at $0.1\si{\astronomicalunit}$, which would not allow the formation of the Kepler-223 system.

From the figure, it appears like the capturing process of this four-planet system into resonance is chaotic and therefore depends very sensitively on both the initial separation and the surface density, where even a small change in the initial distance to the resonance may make the difference between a successful formation of or the failure to create the correct chain.

\subsection{Results with turbulence}\label{sec:turb}
Some of the initial conditions that lead to the successful creation of the Kepler-223 chain are used to investigate the effects of turbulence on the formation of the chain. Three different initial separations were picked, where surface densities ranged between $0.72\,\Sigma_\mathrm{MMSN}$ and $3.93\,\Sigma_\mathrm{MMSN}$. These surface densities were chosen because they allow for most of the successful simulations. Choosing higher surface densities give too high of a migration speed, which is unfavorable for the formation of the resonance chain, while lower surface densities do not offer great constraints for the simulations that include turbulence. The initial conditions that were used are denoted with a blue circle in Fig. \ref{fig:dotsplot}.

\begin{figure*}[htp]
    \centering\includegraphics[width=\textwidth]{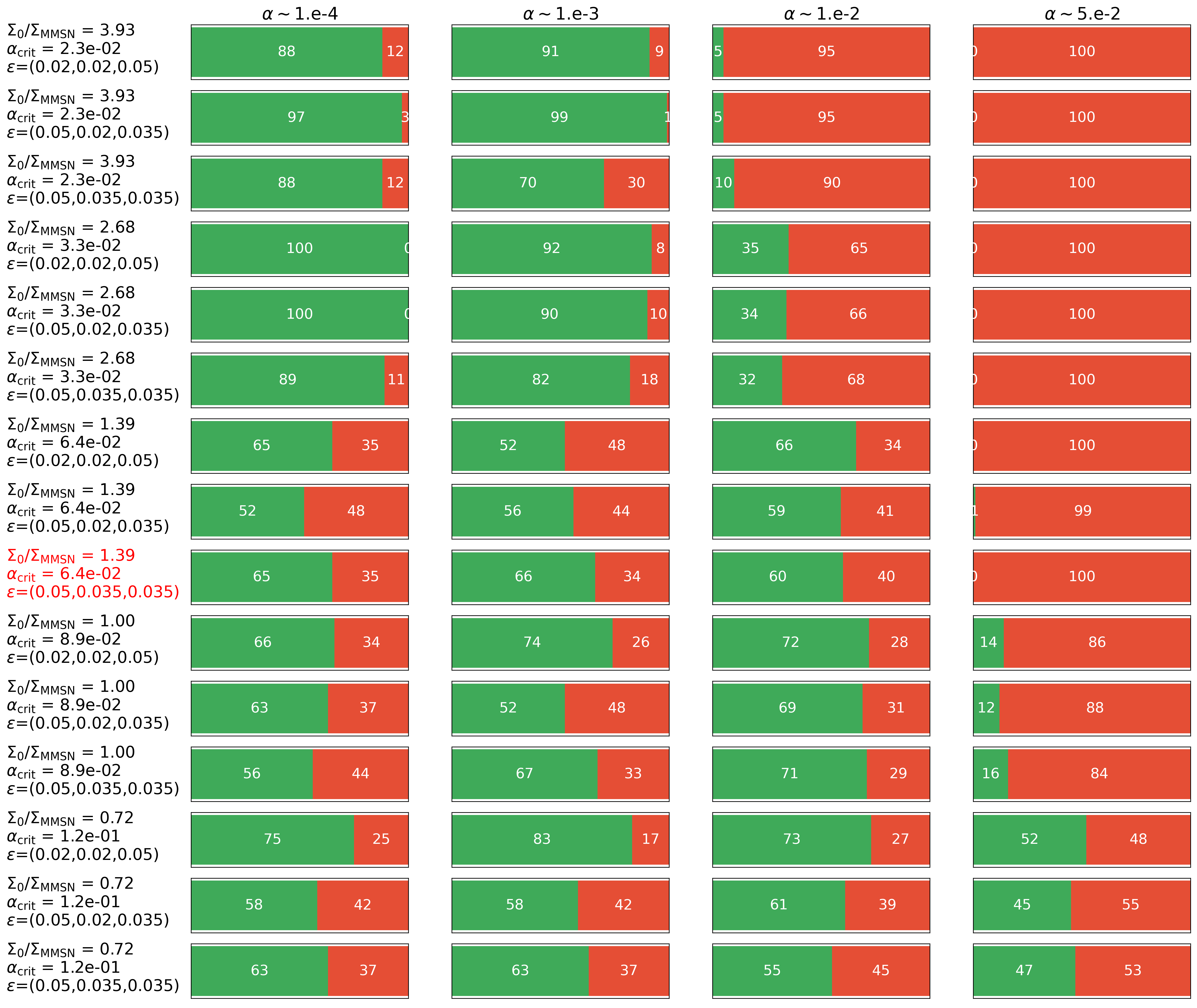}
    \caption{Amount of successful creations of the Kepler-223 resonance chain for various initial conditions, with varying initial separations and surface densities. For each initial setup, four turbulence strengths $\alpha$ were considered. The green bar denotes successful formations. Generally, a larger turbulence strength leads to fewer successful runs. For each setup, there is a critical value $\alpha_\mathrm{crit}$ (cfr. equation \eqref{eq:batygincrit}), for which stronger turbulence significantly hinders the creation of the resonance chain \citep{batygin2017}, which is depicted in the plot. This critical value decreases with increasing gas surface density, which is also the case for the amount of successful runs for increasing surface density at constant $\alpha$ (cfr. Eq. \ref{eq:tauwave}). The setup marked in a red font did not reproduce the Kepler-223 resonance chain without turbulence, but it does not differ significantly from those that did when the effects of the turbulent disk are considered, indicating that turbulence is a necessary ingredient for this particular setup, which remains to be investigated in future work.}
    \label{fig:barsplot}
\end{figure*}

After finding initial conditions that are favorable for the formation of the resonance chain, we investigated the effects of turbulence on its formation to find constraints for the turbulence strength present in the protoplanetary disk during the formation of the chain. We based our analysis on 100 simulations per initial setup. The integration time is not based on physical properties of the system, since the remaining disk lifetime at the point of formation of the chain is not known, and neither is the influence of a possible time variability of $\alpha$. Rather, the integration time was chosen as a compromise between simulation run time and stability of the results; we found that integrating for $\num{e5}$ years achieves this compromise. We also investigated the effects of a longer integration for a particular case, and found that, while some initially stable configuration became unstable, the difference was below 10\%. Therefore, it does not significantly change the results presented in Fig. \ref{fig:barsplot}.

Four different strengths of turbulence were considered, ranging between $\alpha=\num{e-4}$ and $\alpha=\num{5e-2}$. The resulting trend is depicted in Fig. \ref{fig:barsplot}. Several trends are worth pointing out. Firstly, a larger turbulence strength generally leads to less success in forming the Kepler-223 resonance chain. The point at which the formation of the chain becomes significantly improbable lies at lower $\alpha$ for higher surface densities, which fulfills the expectation that the strength of the random kicks be dependent on the product $\Sigma\sqrt{\alpha}$ (cfr. Eqs. \eqref{eq:kickstrength}, \eqref{eq:gammaalpha}). This can be further seen by considering individual columns of the figure, where a larger amount of simulations lead to the formation of the resonance chain when the surface density is lower at constant $\alpha$. However, this trend does not hold for the lower turbulence strengths. This can be seen as an indication that simple analytical trends derived for the two-planet case \citep{batygin2017} are not applicable when the strength of the turbulent kicks is not sufficient to disrupt chains that have already established their resonant configuration (see below), while they can serve as an order-of-magnitude approximation in a case where the turbulence is strong enough to disrupt formed chains. In the former case, disruption has to take place during the capturing process, where simple assumptions do not hold for four planets.

Systems can fail to recreate the Kepler-223 resonance chain for a variety of reasons. Frequent examples include the second planet pair skipping the desired 3:2 resonance and capturing into a 4:3 resonance instead, which can lead to other pairs failing to capture into or stay in the correct resonance. The second planet pair is especially prone to temporary capture, because its desired resonance has the lowest index $k$ and its outer planet is more massive than its inner one \citep{deck2015}. Even in the event of successful capture of the second pair, planets can skip their desired resonance during the initial capturing process of the chain or break out of it later on. Another possible path to failure is created by the brief divergent migration of the third pair, which occurs because its inner planet is more massive than the outer one. If the more massive planet is not captured fast enough into the 3:2 resonance of pair 2, pair 3 can cross the 3:2 resonance from the inside, which is known to cause a rapid excitation of eccentricity. The rapid excitation usually breaks the resonances of other pairs. The frequency at which the third pair crosses the 3:2 resonance is related to the initial separation of the planets. Once a desired resonance is skipped, the planets may either capture into a resonance with higher index $k$, or collide. Rarely, a pair captures into a second-order resonance.

Figure \ref{fig:barsplot} also contains the results of a simulation corresponding to initial conditions that, without considering turbulence, were not favorable for the creation of the Kepler-223 resonance chain. These initial conditions are indicated using a red font. One can see that, when turbulence is considered, the probability of the successful formation of the correct chain does not differ significantly form all other initial conditions considered, even though they allowed for a correct chain in the nonturbulent case. This suggests that turbulence, in this case, is necessary, while even small turbulence strengths are sufficient to bring the probability in line with other similar initial conditions. The reason for this, however, remains to be investigated in future work.

The trend that less simulations lead to the formation of the Kepler-223 system if the turbulence strength is higher is not apparent for lower values for $\alpha$, however it can be clearly seen for higher $\alpha$. One would expect that for an increase in turbulence strength, the amount of simulations that successfully recreate the Kepler-223 resonance chain drops. While this expectation is fulfilled for larger $\alpha$, it is not as clear for lower $\alpha$, where simulations with lower surface densities (i.e. lower turbulence strength) seems to not reproduce Kepler-223 as often as those with a higher surface density do. One can see that, once the strength of turbulence lies beyond a critical value, the disruption of the resonance chain is significantly more efficient than below that value. We propose that this is because turbulence below a certain threshold can only disrupt chain formation while the planets are migrating to the correct semi-major axes, while turbulence beyond that threshold can also disrupt chains efficiently after the planets have already fully captured in resonance. This can also serve as an explanation as to why expected trends for the formation of the resonance chain are not as clear for turbulence below that threshold. In the case of only two planets, \citet{batygin2017} developed an order-of-magnitude criterion for turbulent disruption of formed resonance chains. It is given by
\begin{equation}
    \frac{h}{20}\frac{M}{m_1+m_2}\sqrt{\frac{3\alpha}{f}}\times\left[\frac{\Sigma\langle a\rangle^2}{kM}\sqrt{\frac{\Sigma\langle a\rangle^2}{m_1+m_2}}\right]^{1/3}\gtrsim 1\text{.}\label{eq:batygincrit}
\end{equation}
Here, $m_1$ and $m_2$ are the masses of the planets, $M$ is the stellar mass, $\langle a\rangle$ is the average semi-major axis and $f$ is a factor that accommodates different implementations of the rate of change of the semi-major axis. For our considerations, we apply equation \eqref{eq:batygincrit} to the second planet pair, because it is observed to be in a 3:2 resonance, which has the lowest resonance index of the chain and should therefore be the least stable. The resulting critical value $\alpha_\mathrm{crit}$, obtained by solving \eqref{eq:batygincrit} toward $\alpha$, is also indicated in Fig. \ref{fig:barsplot}. The analytical values match the results from the simulations in order-of-magnitude. In fact, in the four-planet case, values for $\alpha$ that lie slightly below the predicted $\alpha_\mathrm{crit}$ from two-planet considerations lead to the disruption of the formed chain.

\begin{figure*}[htp]
    \begin{subfigure}{\textwidth}
        \centering\includegraphics[width=\textwidth]{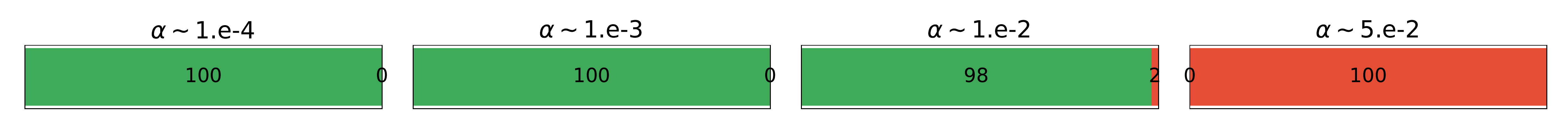}
        \caption{$\Sigma=1.39\Sigma_\mathrm{MMSN}$, $\epsilon=(0.05, 0.02, 0.035)$}
        \label{fig:inplace_1}
    \end{subfigure}\\
    \begin{subfigure}{\textwidth}
        \centering\includegraphics[width=\textwidth]{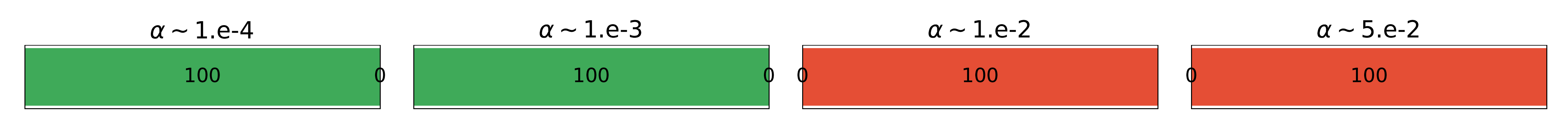}
        \caption{$\Sigma=3.93\Sigma_\mathrm{MMSN}$, $\epsilon=(0.05, 0.035, 0.035)$}
        \label{fig:inplace_2}
    \end{subfigure}
    \caption{Two setups, also represented in Fig. \ref{fig:barsplot}, where turbulence was activated in the simulations only after the formation of the resonance chain. For case (a), equation \eqref{eq:batygincrit} gives $\alpha_\mathrm{crit}=\num{6.4e-2}$ and for case (b),  $\alpha_\mathrm{crit}=\num{2.3e-2}$. The resulting amount of successful simulations as a function of $\alpha$ is shown, which was compared to the expected critical $\alpha_\mathrm{crit}$ value (cfr. Eq. \ref{eq:batygincrit}). The shown results match the theoretical expectations order of magnitude.}
    \label{fig:inplace}
\end{figure*}

Furthermore, the validity of equation \eqref{eq:batygincrit} was tested by considering an initial setup where the four planets are already in resonance. Two initial conditions were considered, the first one with a surface density of $\Sigma=1.39\Sigma_\mathrm{MMSN}$ and $\epsilon=(0.05, 0.02, 0.035)$, and the second one being $\Sigma=3.93\Sigma_\mathrm{MMSN}$ and $\epsilon=(0.05, 0.035, 0.035)$. Equation \eqref{eq:batygincrit} gives $\alpha_\mathrm{crit}=\num{6.4e-2}$ and $\alpha_\mathrm{crit}=\num{2.3e-2}$, respectively. Figure \ref{fig:inplace} shows simulations implementing the aforementioned conditions. It is apparent that the criterion is a good order of magnitude approximation also in the four-planet case, and the behavior of the general case where the planets start outside of the resonances is well-explained by the destruction of formed chains beyond $\alpha_\mathrm{crit}$, where the destruction of formed chains becomes possible at lower values for $\alpha$ than $\alpha_\mathrm{crit}$ because the four-planet case is more sensitive than the two-planet one.

\subsubsection{Constraining the minimum turbulence level}\label{sec:constrain_alpha_lower}
While the results from Fig. \ref{fig:barsplot} can give an upper limit for the strength of disk turbulence during the formation of the Kepler-223 resonance chain, it does not provide a lower limit. The amplitude of libration of the Laplace angles can help in finding a lower limit, as \citet{mills2016} provided the observed amplitudes of these angles.

\begin{figure*}[htp]
    \begin{subfigure}{\textwidth}
        \centering\includegraphics[width=\textwidth]{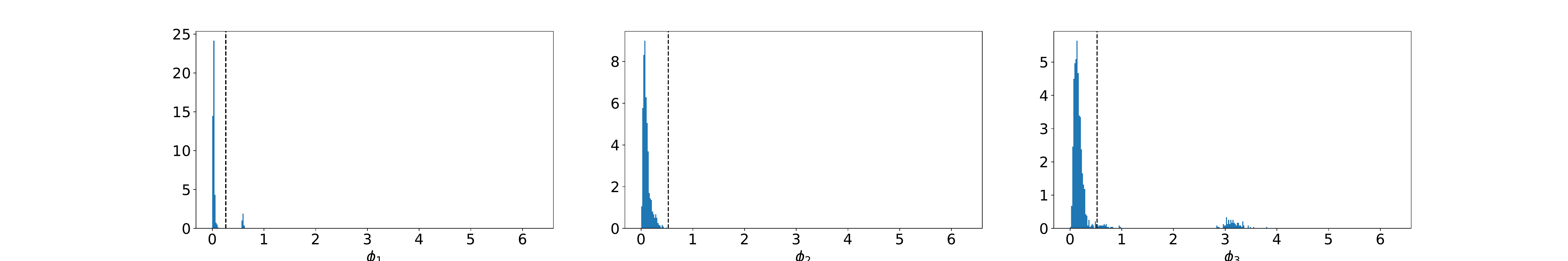}
        \caption{$\alpha=\num{e-4}$}
        \label{fig:amplitudes_1e4}
    \end{subfigure}
    \begin{subfigure}{\textwidth}
        \centering\includegraphics[width=\textwidth]{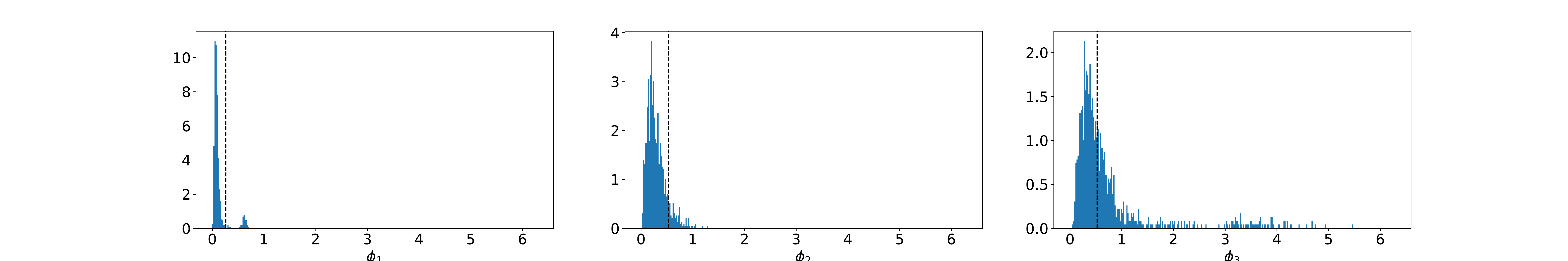}
        \caption{$\alpha=\num{e-3}$}
        \label{fig:amplitudes_1e3}
    \end{subfigure}
    \begin{subfigure}{\textwidth}
        \centering\includegraphics[width=\textwidth]{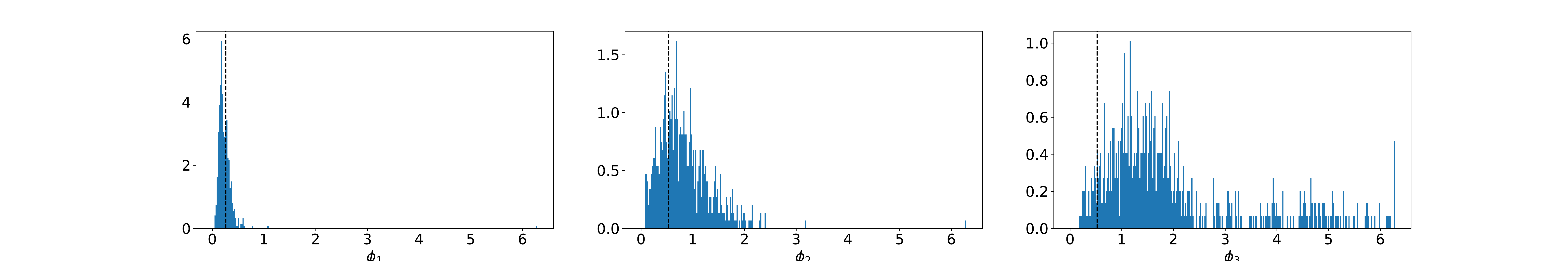}
        \caption{$\alpha=\num{e-2}$}
        \label{fig:amplitudes_1e2}
    \end{subfigure}
    \begin{subfigure}{\textwidth}
        \centering\includegraphics[width=\textwidth]{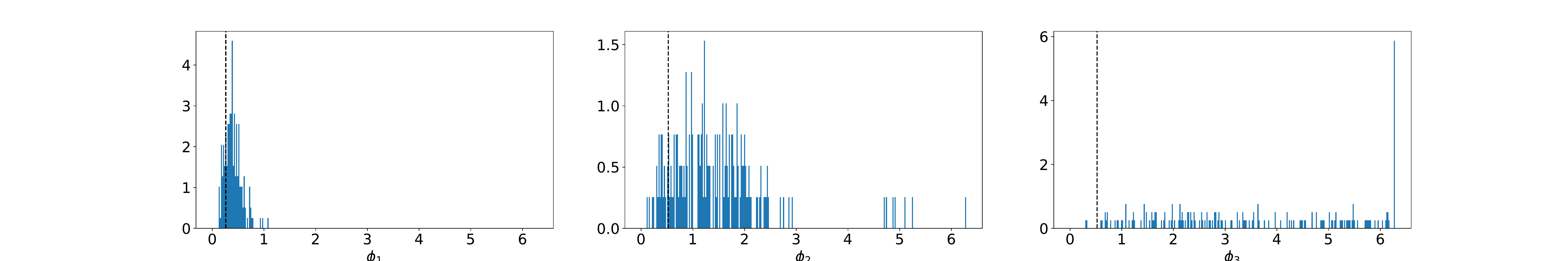}
        \caption{$\alpha=\num{5e-2}$}
        \label{fig:amplitudes_5e2}
    \end{subfigure}
    \caption{The libration amplitudes of the Laplace angles for successful simulations depicted in Fig. \ref{fig:barsplot}, grouped by turbulence strength. The positions of the peaks scale with ${\sim}\sqrt{\alpha}$, as expected due to $D_e\propto\alpha$, which is the same scaling as for the turbulence strength given by equation \eqref{eq:kickstrength} for constant surface density $\Sigma_0$. The expected amplitudes according to \citet{mills2016} are depicted with back dashed lines. The uncertainties of the amplitudes of the Laplace angles are small and therefore are not depicted for our rough value given by those lines. It is apparent that for the lowest turbulence strength $\alpha=\num{e-4}$, the Laplace angle amplitudes are smaller than expected, while for $\num{e-3}\leq\alpha\leq\num{e-2}$, an agreement between the observed amplitudes and the expected ones can be found. For the highest $\alpha=\num{5e-2}$, the amplitudes resulting from the simulations do not match those expected in the case of $\phi_3$, while for $\phi_1$ and $\phi_2$ an agreement can still be found.}
    \label{fig:amplitudes}
\end{figure*}

From an analytical perspective, one would expect the amplitude of the Laplace angles to grow like $\sqrt{D_e}\propto \sqrt{\alpha}\Sigma$, because the diffusion coefficient governs the average evolution of the eccentricity $e$, which in turn gives the evolution of the resonance and Laplace angles. The scaling is the same as for the turbulence strength (cfr. equation \eqref{eq:kickstrength}). For a fixed $\Sigma_0$, this relation can therefore be used to give bounds to $\alpha$. In Fig. \ref{fig:amplitudes}, the libration amplitudes for every Laplace angle, grouped by turbulence strength, are shown for the simulations where the correct chain was formed. The figure demonstrates that the libration amplitudes indeed scale like ${\sim}\sqrt{\alpha}$ for constant surface density, allowing the use of the Laplace angle amplitudes to find a lower limit for $\alpha$. It is shown that for the lowest turbulence strength of $\alpha=\num{e-4}$, the amplitudes of the Laplace angles for the formed chain are smaller than the observed angles. For $\num{e-3}\leq\alpha\leq\num{e-2}$, an agreement between the observed amplitudes and those resulting from the simulations can be found. For the highest strength of $\num{5e-2}$, the amount of successful simulations is low compared to other strengths, and they are mostly limited to low surface densities. With that in mind, the amplitudes for $\phi_3$ do not agree with the simulations, while the amplitudes for $\phi_1$ and $\phi_2$ can match the expected values.

\subsubsection{Eccentricities of the planets}
\begin{figure*}[htp]
    \centering\includegraphics[width=\textwidth]{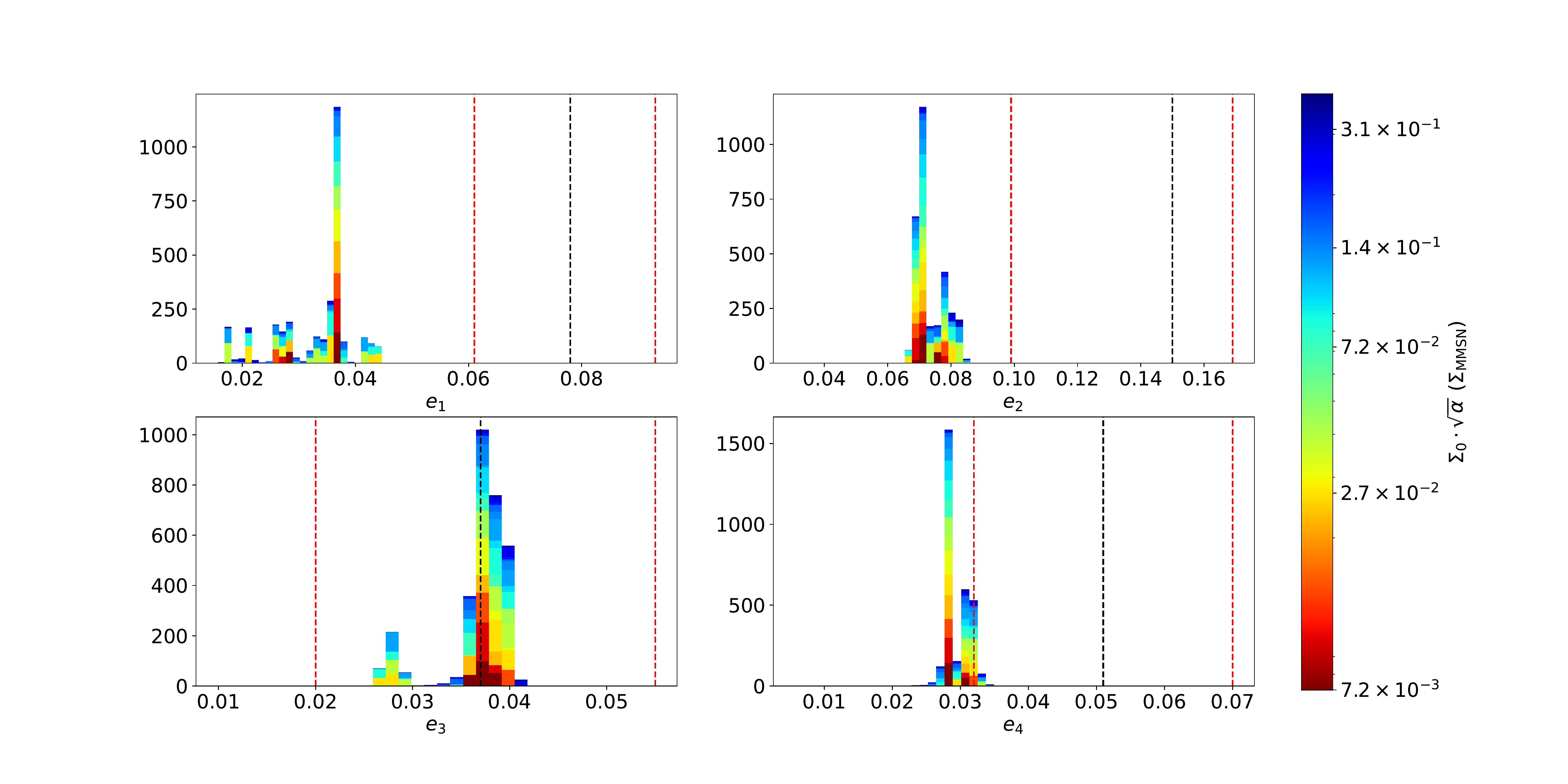}
    \caption{Final averaged planet eccentricities of successfully formed resonant chains, as shown in Fig. \ref{fig:barsplot}. The different turbulence strengths $\Sigma_0\cdot\sqrt{\alpha}$ for the individual simulations are indicated by different colors, as shown by the color bar. The black dashed line denotes the expected eccentricities of the planets \citep{mills2016}, while the red dashed line denotes the $1\sigma$ error. Note that in order to provide better readability, the range of eccentricities depicted on the abscissa is different for the individual plots. Unlike for the two-planet case, a dependence on turbulence strength and initial separation can be found for this case of four planets, resulting in multiple peaks in the distribution. The eccentricities of the two inner most planets do not match the measured values, while for the outer two planets, an agreement can be found. This is probably related to our implementation of the planet trap at the disk's inner edge, which does not capture the entire physical scenario (see Sect. \ref{sec:discussion}).}
    \label{fig:eccentricities}
\end{figure*}

Since \citet{mills2016} also provide the eccentricities of the planets in the Kepler-223 system, these can be compared to the eccentricities that result from the simulations. The comparison between averaged final eccentricity for cases where the correct resonance chain was formed to the eccentricities measured by \citet{mills2016} is depicted in Fig. \ref{fig:eccentricities}. The different turbulence strength values that correspond to the successful simulations are indicated by different colors. In the two-planet case, one would not expect the final eccentricities to depend on the turbulence strength. The reason for this is that random kicks as produced by turbulence only lead to a change around the analytical equilibrium point, the value itself is however independent of $\alpha$ and $\Sigma_0$. The analytical value for the two-planet case can be found as \citep{goldreich2014}
\begin{equation}
    e_\mathrm{eq,1} \approx \left(\frac{1}{2(k+1)}\frac{\tau_e}{\tau_a}\right)^{1/2}=\left(\frac{1}{2(k+1)}K_1\right)^{1/2}\text{,}\label{eq:ecc_eqlib}
\end{equation}
where the subscript 1 denotes an evaluation for the inner planet and $K$ is the K-factor as defined in equation \eqref{eq:kfactor}. This result holds true in the limit of a mass-less inner planet moving outward toward a
massive planet on a fixed, circular orbit (circular restricted three-body problem). A more general description shows that a similar relation also holds for the outer planet's equilibrium eccentricity, $e_\mathrm{eq,2}\sim K_2^{1/2}$ \citep{pichierri2018}. Equations \eqref{eq:ecc_eqlib} and \eqref{eq:kfactor} show that the equilibrium eccentricity is proportional to $h$, and is a function of $k$. The eccentricity distribution shown in Fig. \ref{fig:eccentricities} exhibits multiple peaks, which means that the four-planet case has a dependence on the turbulence strength and initial separation, unlike in the case of only two planets. When comparing our result for the eccentricities with the measured values, we find that there is a mismatch for the two inner-most planets. This could be due to restrictions in our setup, where the inner edge of the protoplanetary disk is modeled by applying high opposing accelerations to the planets close to it, which may not represent the physical situation to high accuracy. In fact, the planet trap could be caused by a sharp drop in surface density, which also affects the eccentricity damping timescale and therefore the equilibrium eccentricity of the planets that are pushed into the trap. By just applying a factor to the acceleration of the semi-major axis, this effect is not considered, resulting in final eccentricities that are lower than they should be. We also note that the eccentricities depend on the aspect ratio, which was fixed at a constant value of $h=0.05$. Increasing the aspect ratio of the disk can therefore in principle lead to a better agreement in the eccentricities.

\section{Discussion}\label{sec:discussion}

In Sect. \ref{sec:constrain_alpha_lower}, we showed that a lower limit for $\alpha$ can be inferred by considering that a certain level of turbulence is needed to reach an excitation of the Laplace angle amplitudes that match those inferred by \citet{mills2016}. While the $\alpha$ values that allow a formation of the Kepler-223 system with the correct Laplace angle libration amplitudes also depend on the disk surface density, a lower limit is given by $\alpha\sim\num{e-3}$. This value is consistent with turbulence values inferred for an MRI "dead-zone" \citep{flock2019}. Even though $\alpha$ has a space dependence across the protoplanetary disk, the assumption of a constant $\alpha$ is not too strong for our approach, considering that the planetessimals that formed the initial embryos resulting in the Kepler-223 planets formed close to each other, as suggested by embryo growth simulations (e.g., \citealt{voelkel2020,voelkel2021}).

To apply the concept of inferring disk parameters from a planetary system after the disk itself has already dissipated, we adopted the prescription by \citet{cresswell2008} for the additional dissipative forces added to the N-body integrator. This description of planetary migration is comparatively simple, therefore allowing to get gross estimates of parameter ranges and giving a good overview over the idea of this method. For a more thorough investigation, more sophisticated prescriptions for planetary migration in protoplanetary disk may be implemented. While this can of course be achieved by integrating hydrodynamical models that are commonly used for investigations of this kind (see e.g., \citealt{bitsch2010,mcnally2019,ataiee2020}), the goal is to do so with the less computation-intensive method of N-body integration including dissipative forces. In this case, adopting for example models from \citet{paardekooper2011,jimenez2017}, a main aspect is the fact that disk torques responsible for migration in fact exhibit a viscosity dependence, which was not considered here. In addition, such models can include more physically accurate models for the inner edge of the protoplanetary disk, where the positive torque acting as a planetary trap is caused by the slope of the density profile, which changes rapidly close to the inner edge. Adopting these considerations in a self-consistent way including stochastic forces as they are caused by the disk turbulence would allow for a more detailed picture of resonant systems, while in principle applying the same arguments as with the simplified approach.

While an N-body approach is of course not as detailed as a full hydrodynamical consideration, it serves as a good way to reduce the large parameter space, because the N-body approach saves a significant amount of computation time compared to a full hydrodynamical approach. In fact, 6000 simulations were run for \num{e5} years of physical time to produce Fig. \ref{fig:barsplot}, which corresponds to about \num{5e6} orbits of the innermost planet Kepler-223 b. One simulation used, on average, 24 CPU hours, so one orbit required ${\sim}\num{4.8e-6}$ CPU hours (${\sim}\SI{17}{\milli\second}$). Under the assumption that a typical hydrodynamical approach takes 4 minutes to integrate one orbit, the methods applied here would have required about \num{3.3e6} CPU hours (${\sim}$38 years) per simulation. The amount of simulations and high integration time that we employed for our considerations would therefore be unfeasible for a hydrodynamical approach.

While planetary systems with long chains of mean-motion resonances are not common due to reasons stated above, Kepler-223 is certainly not the only system with such a peculiar configuration. Another prominent example is the Trappist-1 system, which consists of seven resonant planets. Similarly to the approach taken for Kepler-223, one can infer parameters of the protoplanetary disk in the phase where planetary migration was relevant, even though the disk has already dissipated, by imposing that the disk must allow such a system to form and stay stable during the timescales we observe. This can supplement studies concerning the properties of observable protoplanetary disks, for instance with ALMA \nopagebreak\citep{rafikov2017,dullemond2018,flaherty2017,flaherty2018}. More broadly, any system with a rare orbital configuration, which in turn offers constraints to a protoplanetary disk that would have formed such a rare system, can be used in this way to infer parameter estimates that offer insights on the formation of protoplanetary systems in general.

\section{Conclusions}
   \begin{enumerate}
      \item The observed orbital configuration of the Kepler-223 system can be used to infer information about the protoplanetary disk out of which the system formed. This can be achieved without the need of computationally expensive codes that consider the full hydrodynamical picture, but rather by simply employing an N-body integrator with additional dissipative forces mimicking the relevant planet-disk interactions. However, the approach demonstrated in this work should be seen as a proof-of-concept rather than providing detailed results about the disk parameters that governed the late evolution of the Kepler-223 system. This is because this simple model has a number of limits: No evolution of the protoplanetary disk over time is considered, neither is the space- (or time-) dependence of $\alpha$. Considering these can affect the formation of the chain \citep{ataiee2020} and also the long-term stability (e.g., \citealt{pichierri2018, pichierri2020}). Also, one has to acknowledge that this method probes only the evolutionary phase where migration of planets and disk turbulence are the dominating effects. Nevertheless, our method clearly shows that the dynamical constraint from planetary systems in resonance can be used to constrain the disk parameters during planet formation.
      
      \item For a fixed surface density value $\Sigma_0$, this approach can be used to find constraints for the turbulence parameter $\alpha$, by requiring the formation of the resonance chain under these turbulent conditions. This is done under the simplifying assumption that the planets under investigation can be considered to be close to the resonances that they are observed in today. This can be seen as not too restrictive, considering that it is safe to assume that the planets reached an orbital state close to where they are observed at some point in time. Conversely, under the assumption of a fixed $\alpha$, the required range of surface densities can be inferred, implying whether the planets of Kepler-223 have formed during the early time of high surface density or the late time of lower surface density.
      
      \item For lower values for $\alpha$, it is not straightforward to conclude whether the formation of all aspects of the Kepler-223 orbital architecture can be ensured. This is because for weak turbulence, meaning such that captured resonance chains cannot be disrupted, simple analytical insights cannot be put into use and the impact of the turbulent environment is relevant mostly during the formation stage of the chain. The points at which we recover analytical considerations is reached when also an established chain can be disrupted. The limiting parameter $\alpha_\mathrm{crit}$ can be estimated from two-planet results following \citet{batygin2017}, as described in equation \eqref{eq:batygincrit}.
      
      \item The values of $\alpha$ that were considered in this work are in the range of $\alpha=\num{e-4}$ to $\num{4e-2}$, which is the range found by studies of disk observable with ALMA \citep{rafikov2017,dullemond2018,flaherty2017,flaherty2018}. These observations are also consistent with hydrodynamical and MRI simulations (e.g., \citealt{turner2014}). When just considering the assembly of the correct chain of mean-motion resonances, it is not possible to give a lower limit for $\alpha$ using the methodology employed here. However, since also the amplitudes of the Laplace angles are known for the Kepler-223 system \citep{mills2016}, an additional constraint can be provided. As shown in Fig. \ref{fig:amplitudes}, a certain level of turbulence is needed to reproduce the correct amplitudes of the Laplace angles, which is about 15° for $\phi_1$ and 30° for $\phi_2$ and $\phi_3$, peak-to-peak. As a lower limit for $\alpha$, the consideration of the Laplace angle amplitudes gives $\alpha\sim\num{e-3}$, which is consistent with values typically considered for an MRI "dead-zone" \citep{flock2019}. Our measurement of viscosity comes with the uncertainty of an unknown value for the surface density, since our simulations allow us to make statements about $\Sigma_0\sqrt{\alpha}$. For example, a higher turbulent viscosity can be consistent with our results if we adopt a lower surface density. It was also found that even if the correct MMR chain is formed, the system can still fail to reproduce the correct amplitudes for the Laplace angles and can therefore not be considered to have correctly reproduced the observations, mainly important at high $\alpha$.
      
      \item While the described methods can be used to find rough constraints for disk parameters during the later stages of planet formation and the assembly of planetary systems, it has shortcomings, which can be seen by the example of the final planet eccentricities as depicted in Fig. \ref{fig:eccentricities}. The eccentricities of the inner two planets as measured by \citet{mills2016} were not reproduced here. However, this does not invalidate our method, because on the one hand, a global mismatch of the eccentricities can be fixed by modifying the aspect ratio prescription, where an increase could lead to a shift in eccentricities that provides a better match with the observed values \citep{goldreich2014,pichierri2018}. On the other hand, we propose as a reason for the presented mismatch the model of the inner edge of the protoplanetary disk, which is strongly simplified and does not accurately represent all physical properties. While the focus of this investigation was on the methodology itself, this issue could be fixed by employing more physically accurate models (e.g., \citealt{izidoro2017,izidoro2019}).
   \end{enumerate}
   Our method clearly shows that the currently observed system architectures hold imprints to their formation history. Especially systems with multiple planets in resonance can help to constrain disk parameters. This is especially important to also probe different environments, like in the Trappist-1 system, orbiting an M-dwarf, where constraints from disk observations are even rarer.

\begin{acknowledgements}
L.-A.H, G.P. and B.B. thank the European Research Council (ERC Starting Grant 757448-PAMDORA) for their financial support.
\end{acknowledgements}

\bibliography{references}

\begin{appendix}

\section{Diffusion Coefficient Fit for the Eccentricity Evolution}\label{sec:de_fit}
\begin{figure}[htp]
    \centering\includegraphics[width=.5\textwidth]{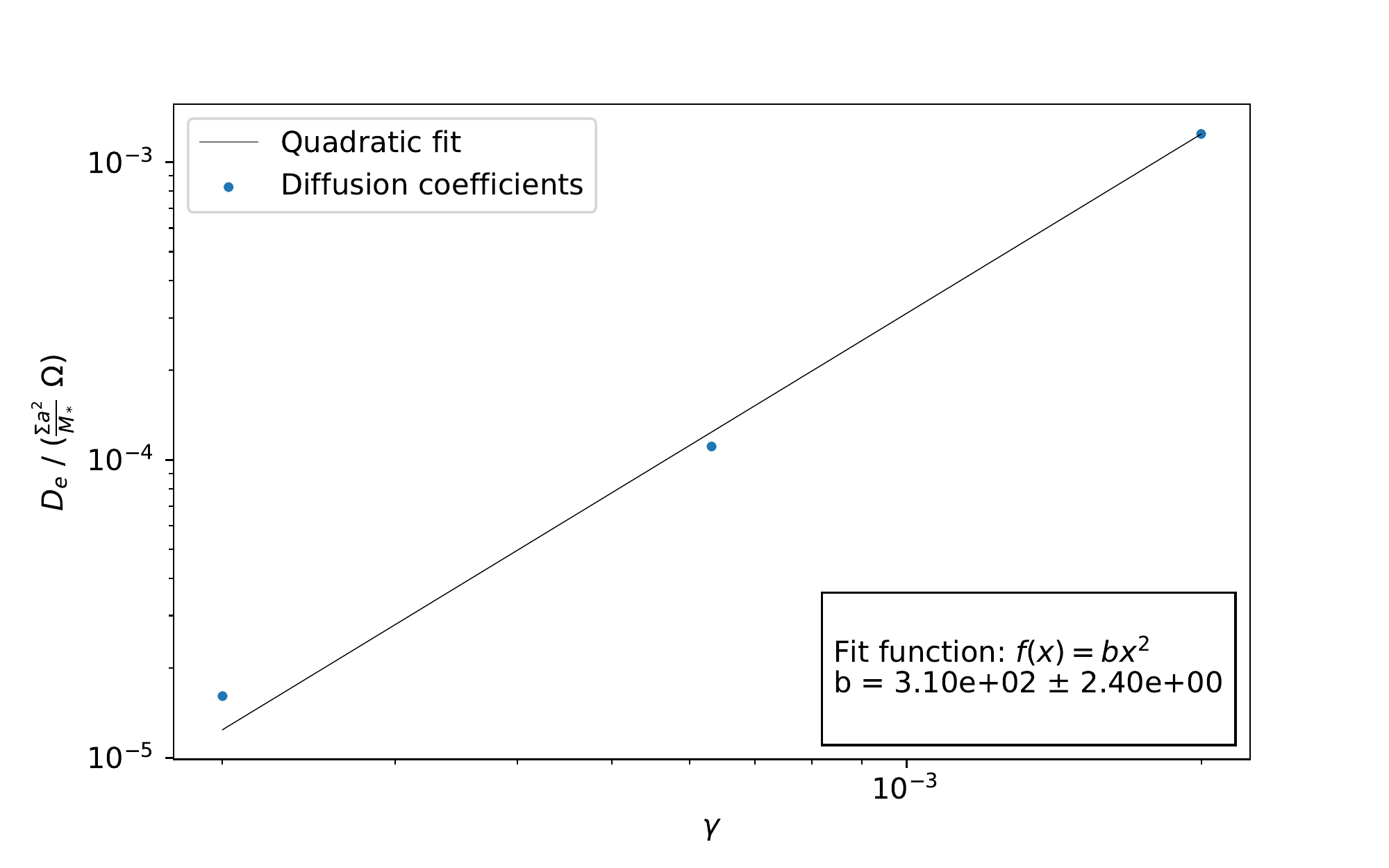}
    \caption{Diffusion coefficient $D_e$ that are realized by three different turbulence strength parameters $\gamma$, in simulations with the setup used throughout this work. A quadratic fit was used to find the pre-factor that describes the order-of-magnitude relation as described in equation \eqref{eq:de_gamma}.}
    \label{fig:de_gamma}
\end{figure}

To derive the pre-factor of equation \eqref{eq:de_gamma}, the diffusion coefficient was derived for different $\gamma$, by computing the standard deviation between 100 simulations for a fixed time, and fitting a square-root dependence. For those simulations, migration was disabled. Figure \ref{fig:de_gamma} shows the resulting diffusion coefficient, where another fit lead to the pre-factor of $\num{3e2}=\num{3e-4}\left(\frac{1}{\num{e-3}}\right)^2$ for the relation between the turbulence strength parameter $\gamma$ and the diffusion coefficient of eccentricity, $D_e$, as seen in equation \eqref{eq:de_gamma}.

\section{Simulation Parameters}
This appendix presents all constant parameters employed for the prescription of migration \citep{cresswell2008} and stochastic forces mimicking turbulence \citep{ogihara2007}.\\

{%
\begin{tabular}{lc}
\hline\hline
Parameter & Value \\
\hline
    $n_\mathrm{modes}$ & 500\\
    $r_\mathrm{in}$ & 0.073 AU \\
    $r_\mathrm{out}$ & 10 AU\\
    $h$ & 0.05 \\
    $r_0$ & 1 AU\\
    $\alpha_\Sigma$ & 1.5\\
\hline
\end{tabular}%
}

\end{appendix}

\end{document}